\documentclass[twocolumn,aps,prb,superscriptaddress,floatfix]{revtex4-2}

\usepackage{graphicx}
\usepackage{dcolumn}
\usepackage{bm}
\usepackage{amsmath}
\usepackage{amssymb}
\usepackage{amsmath}
\usepackage{bbold}
\usepackage[T1]{fontenc}
\usepackage[unicode=true,pdfusetitle,
 bookmarks=true,bookmarksnumbered=false,bookmarksopen=false,
 breaklinks=false,pdfborder={0 0 1},backref=false,colorlinks=false]
 {hyperref}
\hypersetup{
 colorlinks,linkcolor=blue,citecolor=blue,urlcolor=blue}

\usepackage{soul}
\usepackage{xcolor}

\newcommand{\jose}[1]{}
{}
\newcommand{\fig}[1]{Fig. #1}
\newcommand{\subfig}[2]{\fig{#1}(#2)}

\def\ka{\alpha}
\def\kb{\beta}

\begin{document}


\title{Orbital Hall effect and topology on a two-dimensional triangular lattice: from bulk to edge }
\author{Anderson L. R. Barbosa}
\email{anderson.barbosa@ufrpe.br}
\affiliation{Departamento de F\'{\i}sica, Universidade Federal Rural de Pernambuco, 52171-900, Recife, PE, Brazil}
\author{Luis M. Canonico}
\affiliation{Catalan Institute of Nanoscience and Nanotechnology (ICN2), CSIC and BIST, Campus UAB, Bellaterra, 08193 Barcelona, Spain}

\author{Jose H. Garc\'ia}
\affiliation{Catalan Institute of Nanoscience and Nanotechnology (ICN2), CSIC and BIST, Campus UAB, Bellaterra, 08193 Barcelona, Spain}
\author{Tatiana G. Rappoport}
 \email{tgrappoport@fisica.uminho.pt}
 \affiliation{Centro de Física das Universidade do Minho e do Porto (CFUMUP) e Departamento de Física, Universidade do Minho, P-4710-057 Braga, Portugal}
\affiliation{Instituto de Física, Universidade Federal do Rio de Janeiro\\
C.P. 68528, 21941-972 Rio de Janeiro RJ, Brazil}%

\date{\today}

\begin{abstract}

We investigate a generalized multi-orbital tight-binding model on a triangular lattice, a system prevalent in a wide range of two-dimensional materials, and particularly relevant for simulating transition metal dichalcogenide monolayers. We show that the interplay between spin-orbit coupling and different symmetry-breaking mechanisms leads to the emergence of four distinct topological phases [Eck, P., \textit{et al.}, Phys. Rev. B, 107 (11), 115130 (2023)]. Remarkably, this interplay also triggers the orbital Hall effect with distinguished characteristics. Furthermore, by employing the Landauer-B\"uttiker formula, we establish that in the orbital Hall insulating phase, the orbital angular momentum is carried by edge states present in nanoribbons with specific terminations.  We also show that, as expected, they do not have topological protection against the disorder of the edge states belonging to a first-order topological insulator.

\end{abstract}

\maketitle

The past few decades have seen a growing interest in manipulating the quantum degrees of freedom for developing new quantum devices. The emergence of two-dimensional materials and the tunability of their properties added to their capability to sustain highly crystalline interfaces reignited the excitement in spintronics and provided a platform for tackling long-standing challenges in the field while paving the way for groundbreaking technological applications like spin-orbit-based memory systems and ultra-compact devices \cite{dieny2020opportunities,NatNanoSierra2021,YangNature2022}. The tunability of the properties of the 2D materials also enabled the control of the spin-orbit coupling, which is essential for the electrical manipulation of the spins. Nonetheless, due to its electrostatic origin, this interaction tends to appear on heavy metals, which are scarce, and their usage poses environmental risks.

Recently, researchers have turned their attention to the manipulation of the orbital degrees of freedom and exploiting novel phenomena such as the orbital Hall effect \cite{OHEorigins-1,OrbitalTexture,Salemi2022,Negative-intrinsic-OHE,Orientational-dependence-OHE,Sala2022,fonseca2023orbital, OHE_Bhowal-Vignale,Pezo2022,PezoDisorder,Busch2023} and the charge-to-orbital conversion via orbital Rashba-Edelstein effect \cite{Park2012,Johansson2021,Ding2022,ElHamdi2023}. In analogy with their spin counterparts, these effects enable the electrical manipulation of the orbital degrees of freedom without a mediating interaction like the spin-orbit coupling, opening the possibility of using light elements in novel sustainable devices \cite{Rappoport2023} and reigniting the interest in orbitronics \cite{OHEBernevig,Go-Review}. Novel experimental studies have demonstrated that the orbital Hall effect (OHE) can arise in three-dimensional (3D) systems even if the orbital angular momentum of the system vanishes in equilibrium \cite{Choi2023,Lyalin2023} and the orbital currents may have very large decay lengths \cite{Hayashi2023,Seifert2023}. 

Furthermore, it has been shown that the OHE can induce magnetization dynamics via the orbital torque in magnetic hetero-bilayers \cite{go2020orbital,Orbital-torque-1,orbital-torque-magnetic-bilayers-EXP,Fukunaga2023}. Also, the reciprocal version of this effect, the orbital pumping, where an oscillating magnetic moment creates an orbital current,  was recently confirmed \cite{go2023orbital,hayashi2023observation}. In the case of two-dimensional (2D) materials, the interplay between their band structure, orbital degrees of freedom, and the presence of sizable Berry curvature pockets in reciprocal space has garnered significant attention. Theoretical works have shown that, like their 3D counterparts, multi-orbital 2D materials can host orbital textures that trigger sizable OHE \cite{Us1}. Additionally, several theoretical works have predicted that monolayers \cite{Us2,OHE_Bhowal_1,pezo2022orbital} and bilayers \cite{Us3,Cysne-Bhowal-Vignale-Rappoport} of transition metal dichalcogenides (TMDs) exhibit OHE within their energy gap and are characterized by an orbital Chern number. Furthermore, these materials exhibit orbital currents that flow through their edge states, which were previously considered trivial from the $\mathbb{Z}_2$ perspective \cite{Edgestates-TMD-1,Edgestates-TMD-2}.

Recently, a theoretical work by Costa \textit{et al.} \cite{Costa2023} demonstrated the connection between the orbital Hall insulating phase of certain transition metal dichalcogenides (TMD) monolayers and their high-order topological insulating phases \cite{HOTI-Rotation-2,Monalyer-TMD-corner-states,Bilayer-TMD-corner-states}, hinting a connection between the non-topological edge states and the orbital Hall conductivity plateaus \cite{Us2,Us3,Cysne-Bhowal-Vignale-Rappoport,OHE_Bhowal_1}, and evinced the existence of the OHE in centrosymmetric systems. They showed that the accumulation of orbital angular momentum can occur at the edges of ribbons of these materials due to the presence of metallic states that depend on the ribbon orientation. However, despite this array of theoretical predictions, direct experimental evidence of the transport and manipulation of the atomic orbital angular momentum remains elusive.


The development of orbitronic devices based on 2D materials requires a robust criterion to identify orbital-Hall-capable materials. Recent studies by Han \textit{et al.} have taken a significant step toward this goal \cite{Han2022,Han2023}. The authors conducted a detailed analysis of the microscopic origin of the orbital textures and uncovered the interplay between crystalline and orbital symmetries that promote hybridizations leading to the emergence of these textures. From another perspective, Eck \textit{et al.} \cite{Eck2022} recently presented the conditions that enable the appearance of 2D high-order topological insulators (HOTI) on the triangular lattice from the interplay between the crystalline and orbital symmetries. Our current study further explores the correlation between orbital hybridization, orbital symmetries, and the OHE to gain insights into its connection with various topological phases.


In this work, building on the previous work of Eck \textit{et al.}, we explored how symmetry reduction can trigger different spin and orbital topological phases. To this end, we focus on the $d_{z^2}$, $d_{xy}$, and $d_{x^2-y^2}$ orbitals within the $d$-subshell, residing on a triangular lattice; this is a well-established model for characterizing TMD monolayers, where the symmetries can be controlled by removing specific hopping elements. This flexibility allowed us to explore various topological phases, which are subsequently probed through quantum transport calculations of the orbital and spin conductivities using the Kubo and Landauer-B\"uttiker formalisms. Combining these methods, we identified the distinctive orbital-Hall conductivity plateaus within the insulating gap of high-order topological insulators \cite{HOTI-Rotation-2,Monalyer-TMD-corner-states,Bilayer-TMD-corner-states}, determined whether the metallic edge states emerge within the energy gap for specific ribbon orientations, and studied the impact of the disorder on the transport of orbital angular momentum in device geometries.

\section{The model}

 We begin with the three-band tight-binding model, which is commonly used to describe the low-energy behavior of H-TMDs. In this model, a triangular lattice accommodates the $d_{z^2}$, $d_{xy}$, and $d_{x^2-y^2}$ orbitals. To simplify the analysis, we adopt the same tight-binding parameters as in Ref. \onlinecite{ThreeBandTMD} for MoS$_2$, but with a spin-orbit coupling strength that is $15$ times larger than the original value. This allows us to transition between different phases of the system by simply removing some of the original hopping elements, which modify the symmetries of the model without altering the values of the other parameters. Using the basis $\{d_{z^{2}},d_{xy},d_{x^{2}-y^{2}}\}$, the Hamiltonian of the system can be written as $\mathcal{H}=\sigma_0\bigotimes H_0+H_{SOC}$, where

\begin{equation}
H_0=\begin{bmatrix}h_{0} & h_{1} & h_{2}\\
h_{1}^{*} & h_{11} & h_{12}\\
h_{2}^{*} & h_{12}^{*} & h_{22}
\end{bmatrix},
\label{eqn:Hamiltonian3Bands}
\end{equation}
with
\begin{align}
&h_{0}=2t_{0}(\cos2\alpha+2\cos\alpha\cos\beta)+\epsilon_{1}, \nonumber\\
&h_{1}=-2\sqrt{3}t_{2}\sin\alpha\sin\beta+2it_{1}(\sin2\alpha+\sin\alpha\cos\beta)\nonumber\\
&h_{2}=2t_{2}(\cos2\alpha-\cos\alpha\cos\beta)+2\sqrt{3}it_{1}\cos\alpha\sin\beta,\nonumber \\
&h_{11}=2t_{11}\cos2\alpha+(t_{11}+3t_{22})\cos\alpha\cos\beta+\epsilon_{2},\nonumber\\
&h_{22}=2t_{22}\cos2\alpha+(3t_{11}+t_{22})\cos\alpha\cos\beta+\epsilon_{2},\nonumber\label{eq:H2222}\\
&h_{12}=\sqrt{3}(t_{22}-t_{11})\sin\alpha\sin\beta
+4it_{12}\sin\alpha(\cos\alpha-\cos\beta), \nonumber\\
&\text{and }\alpha=\frac{1}{2}k_{x}a \mbox{ and } \beta=\frac{\sqrt{3}}{2}k_{y}a.
\end{align}
 This is a symmetry-based tight-binding Hamiltonian that does not originate from a Slater-Koster approach, as the hopping elements between the $d$ orbitals are mediated by the $p$ orbitals of the chalcogenes in 2H-TMDs.  By examining the TMD Hamiltonian, it becomes apparent that, apart from the term proportional to $t_2$, Eq. \ref{eqn:Hamiltonian3Bands} is formally equivalent to a Slater-Koster Hamiltonian utilizing $p_x$, $p_y$, and $p_z$ orbitals in a two-dimensional triangular lattice where the inversion and mirror symmetries of the lattice are broken (see Appendix A).

In this map between the two Hamiltonians, $t_0$, $t_{11}$ and $t_{22}$ are Slater-Koster hopping elements for $\sigma$ and $\pi$ bonds for $p_z$, $p_x$ and $p_y$ orbitals that respect all the symmetries of a two-dimensional triangular lattice. The terms proportional to $t_0$, $t_{11}$ and $t_{22}$ describe the nearest-neighbor hopping in a triangular lattice of layer group  $p6/mmm$, which is generated by a six-fold rotation, three vertical reflection planes $\sigma_v$, three diagonal reflection planes $\sigma_d$ and one horizontal reflection plane $\sigma_h$ \cite{Eck2022}. 

If inversion symmetry is broken, the six-fold rotation ($C_6$) reduces to $C_3$, and the Hamiltonian presents the term proportional to $t_{12}$. On the other hand, if only the mirror symmetry with respect to the plane is broken, it is possible to have hybridization between  $p_z$ and the in-plane orbitals $p_x$ and $p_y$, which gives rise to the term proportional to $t_1$. The extra term proportional to $t_2$  is the only one that cannot be mapped into a Slater-Koster Hamiltonian. It is important to notice, however, that although we can formally map one Hamiltonian into another, the meaning of the different terms is not the same for the $d$ orbitals. For instance, $t_1$ hybridizes orbitals belonging to different irreducible representations, as in the case of the $p$ orbitals, but does not break the same mirror symmetry, which is preserved in 2H TMDs monolayers.

The spin-orbit coupling is written as
\begin{equation}
H_{\rm SOC}=\sigma_z\frac{\lambda}{2}\begin{bmatrix}0 & 0 & 0\\
0 & 0 & 2i\\
0 & -2i & 0
\end{bmatrix}.
\end{equation}

\begin{figure}[h]
\centering
\includegraphics[width=0.98\linewidth]{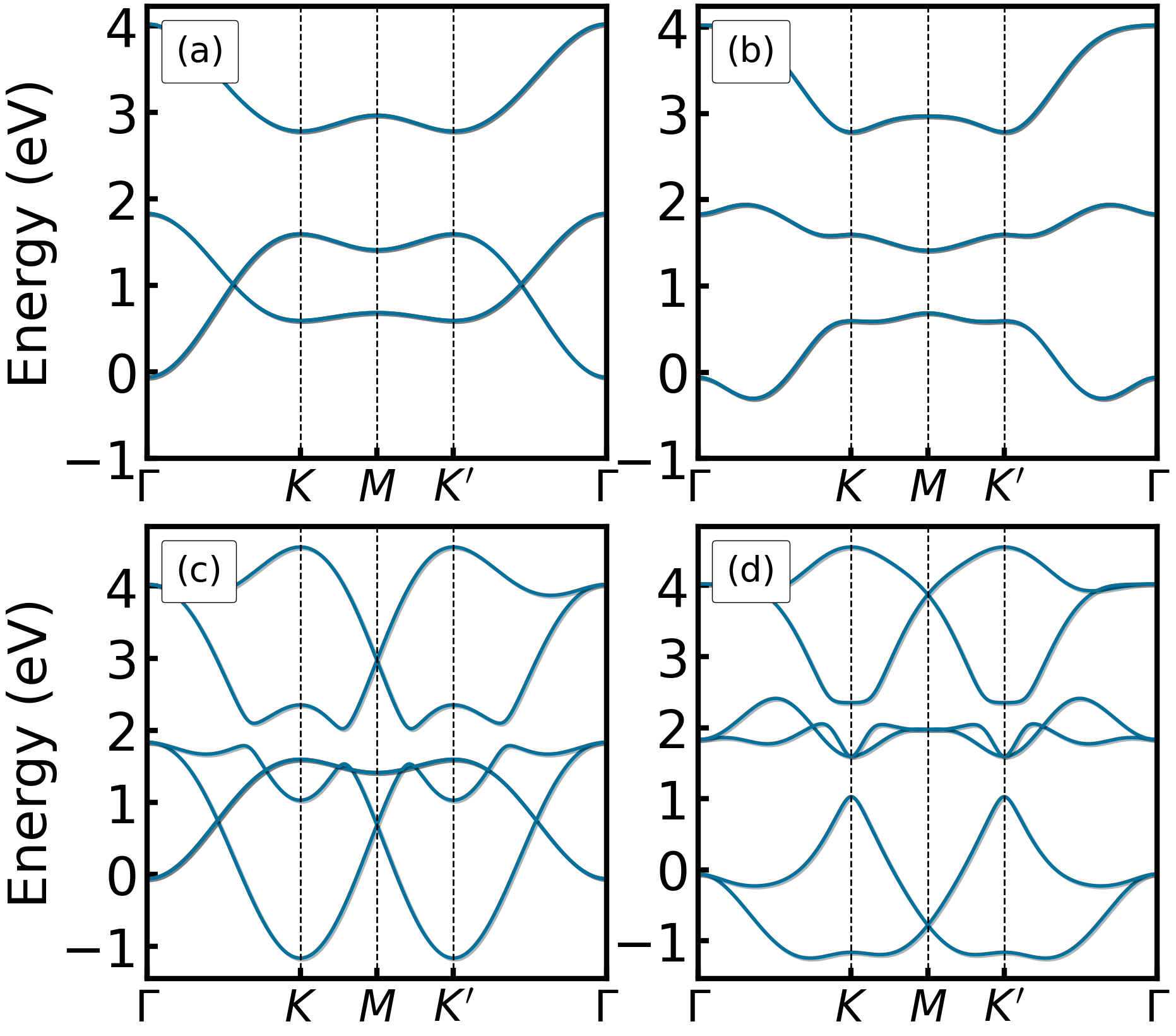}
\caption{Bulk energy bands of the four topological phases of the tree bands model for various terms in the Hamiltonian. Phase I (a): $t_{1}=t_{2}=t_{12}=0$.  Phase II (b): $t_{2}=t_{12}=0,t_1\neq0$. Phase III (c): $t_1= t_2 = 0, t_{12}\neq 0$. Phase IV (d): $t_1\neq t_2\neq t_{12}\neq 0$.}\label{fig:fig1}
\end{figure}

\noindent Here, $\lambda$ is the strength of the spin-orbit coupling, and $\sigma_0$ and $\sigma_z$ are the $2\times2$ identity matrix and the $z$ Pauli matrices, respectively. To investigate the connection between topology and the orbital Hall plateau, we analyze four different phases. To consider the different phases without modifying the spin-orbit coupling or the strength of the other hoppings, we take $\lambda$ as fifteen times stronger than the SOC of MoS$_2$. The band structures of the four phases are shown in figure \ref{fig:fig1}. For phase I, $t_1=t_2=t_{12}=0$, which, in the case of $p$ orbitals, is equivalent to a system where all the symmetries of the triangular lattices are kept. \subfig{\ref{fig:fig1}}{a} shows the band structure of the phase. For this case, the subspace for $m_l=0$ is separated from the subspace where $m_l=\pm l$. Because of the inversion symmetry, all bands are spin degenerate.  The system has a lower band with $m_l=0$.  The two other bands that belong to the $m_l\pm l$ sub-space are split by the spin-orbit coupling.

For phase II, in the case of $p$ orbitals, it indicates that the mirror symmetry is broken with $t_1=t_2=0$ and $t_{12}\neq 0$, while preserving the inversion symmetry, and consequently, the spin-degenerate energy bands. However, there is a finite hybridization between the bands belonging to the $m_l=\pm l$ subspace and the one for $m_l=0$. As a result, the band structure of \subfig{\ref{fig:fig1}}{b} presents a gap at the band crossing of Fig. \ref{fig:fig1}(a).

In the case of phase III, the subspaces of $m_l=\pm l$ and  $m_l=0$ are decoupled but the inversion symmetry is broken. As a result, the bands in \subfig{\ref{fig:fig1}}{c} can present spin and orbital angular momentum textures. This breaks the spin degeneracy of the bands belonging to the $m_l=\pm l$ subspace. 
Phase IV presents all the parameters of a three-band model for MoS$_2$ so that all hopping of the Hamiltonian are finite and a much stronger spin-orbit coupling that produces a large splitting of the conduction band states, as seen in \subfig{\ref{fig:fig1}}{d} but still keeps the spin and orbital character of the TMDs in the vicinity of the band gap.

\section{Topology and the spin and orbital angular momentum transport}

To explore the topology of the different phases, we use a two-fold approach: we calculate the band structure of a nanoribbon and analyze the edge states for the four different phases, while comparing it with the spin and orbital Hall conductivities of the bulk system. For the computation of the spin and orbital Hall conductivities, we used the Kubo-Bastin formula \cite{bastin1971quantum,Garcia2015,Garcia2016}:
\begin{multline}
    \sigma_{\alpha\beta}^{\mathcal{O}}(\mu,T) =  ie \hbar\int_{-\infty}^{\infty}d\varepsilon f(\varepsilon,\mu,T) \times \\Im \left(Tr\langle J_\alpha^{\mathcal{O}}\delta(\varepsilon - H)v_\beta\partial_{\varepsilon}G^{+}(\varepsilon) \rangle \right),
\end{multline}
where $J_\alpha^{\mathcal{O}}=\frac{1}{2\Omega}\lbrace v_{\alpha},\mathcal{O}\rbrace$ is the angular momentum current density operator with $\mathcal{O}$ being either the $S_z$ spin operator or the $L_z$ orbital angular momentum operators, where the later is evaluated within the atom-centered approximation, $v_{\alpha}$ is the $\alpha$ component of the velocity operator in the  $v_\alpha=-\frac{i}{\hbar}\left[ H, R_{\alpha}\right]$ with $R_\alpha$ being the projection of along the $\alpha$ direction of the position operator, $H$ is the Hamiltonian of the system and $\Omega$ the area of the sample. $G^{+}(\varepsilon)$ and $\delta(\varepsilon - H)$ are the retarded Green's and spectral functions, respectively. These are approximated using the kernel-polynomial method \cite{kite,fan2021linear} as implemented in the \textsc{LSQUANT} toolkit \cite{lsquant_page}. For all of our transport calculations, we used $1024$ Chebyshev moments with the Jackson kernel to obtain an energy resolution of $\delta \approx 3$ meV, while assuming periodicity allowed us to consider a grid of $512\times512$ points in reciprocal space.

\begin{figure}[t]
    \centering
    \includegraphics[width=0.98\linewidth]{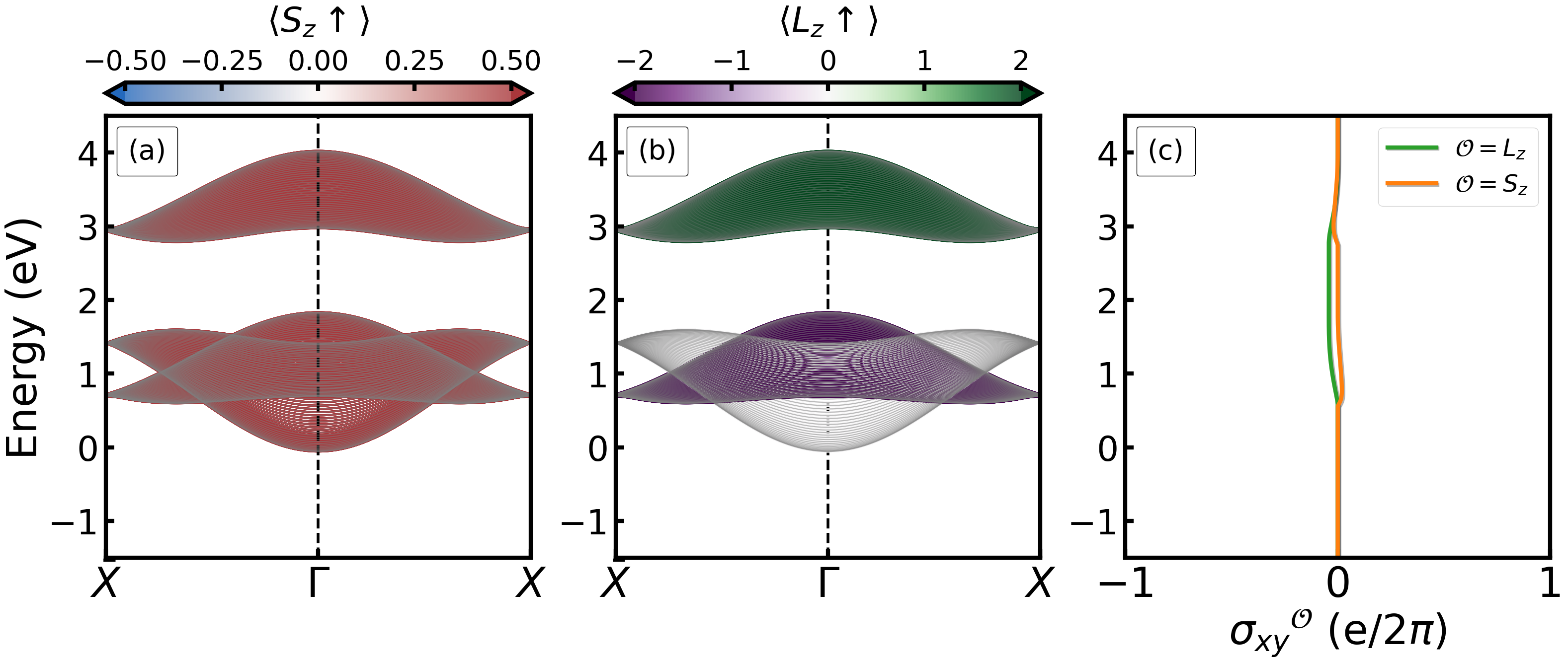}
    \caption{Spin (a) and orbital (b) character of the spin up energy bands of a nanoribbon with breadth $W=50\vec{a}_2$ for a system in the phase I. (c) Spin (orange) and orbital (green) Hall conductivities for a system in phase I. }
    \label{fig:fig2}
\end{figure}

Starting with phase I, \subfig{\ref{fig:fig2}}{a} and \subfig{\ref{fig:fig2}}{b} show the band structure of a nanoribbon with the color code of panels \subfig{\ref{fig:fig2}}{a} and panels \subfig{\ref{fig:fig2}}{b} indicating the $S_z$ and $L_z$  projections respectively. Because of the spin degeneracy, we project the spin-up states. The band structure of the nanoribbon does not have in-gap edge states, signaling that the system lies in a topologically trivial state. Additionally, from the spin and orbital projection of the energy states of the nanoribbon, it is clear that the vanishing contribution from hoppings $t_1,t_2$ and $t_{12}$ forbids the hybridization between the orbitals with $d_{z^2}$, and $d_{x^2-y^2},d_{xy}$ and makes the system inversion symmetric. Consequently, the SOC enforces the energy states formed by linear combinations of orbitals $d_{x^2-y^2}$ and $d_{xy}$ to have a well-defined atomic angular momentum character. Thus making the system a trivial SOC insulator, as the null orbital and spin Hall conductivities presented in \subfig{\ref{fig:fig2}}{c} confirm.

\begin{figure}[h]
    \centering
    \includegraphics[width=0.98\linewidth]{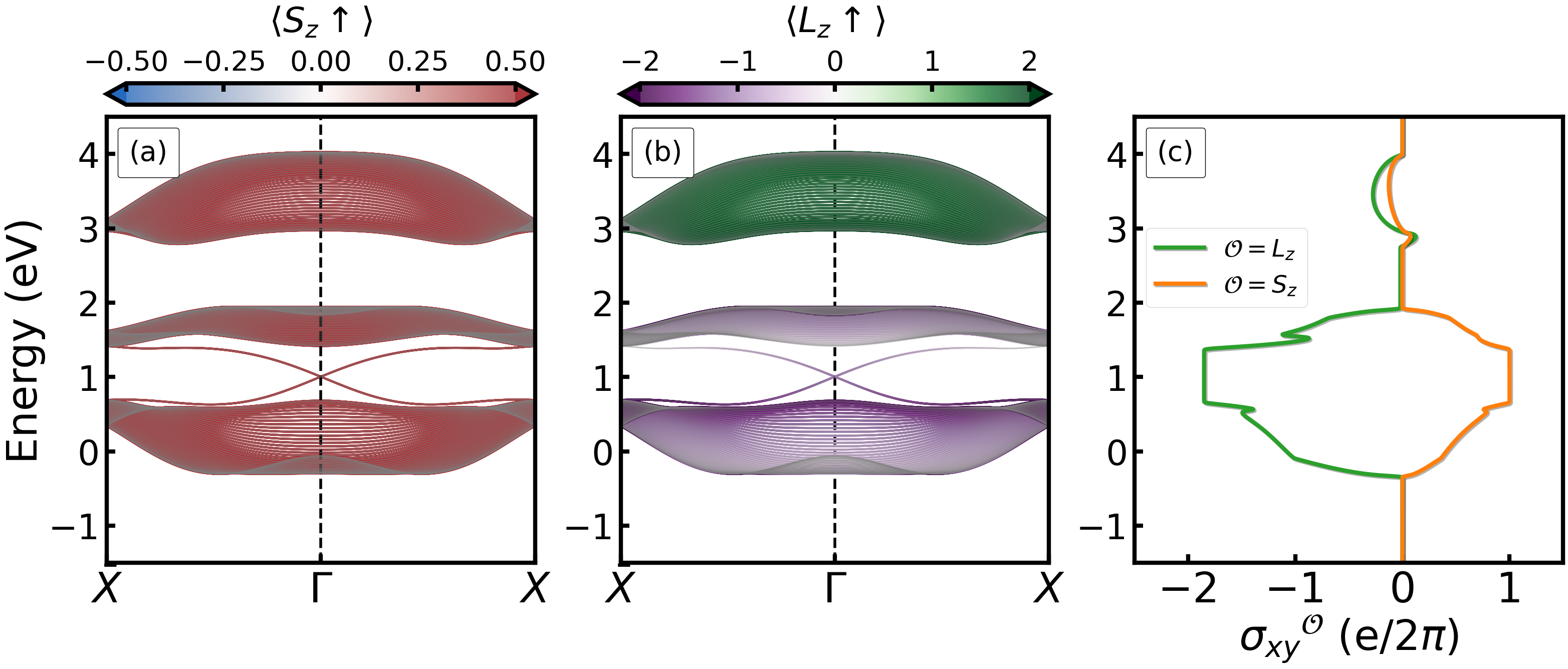}
    \caption{Spin (a) and orbital (b) character of the spin-up energy bands of a nanoribbon with breadth $W=50\vec{a}_2$ for a system in phase II. (c) Spin (orange) and orbital (green) Hall conductivities for a system in phase II. }
    \label{fig:fig3}
\end{figure}

 For phase II, the band structure of the nanoribbon in \subfig{\ref{fig:fig3}}{a} and \subfig{\ref{fig:fig3}}{b} show that after enabling the hybridization between the orbitals $d_{z^2}$, $d_{xy}$, and $d_{x^2-y^2}$ through $t_1$ while maintaining $t_2=t_{12}=0$, the system exhibits inversion symmetry with two pairs of in-gap spin-polarized edge states. Interestingly, the orbital-angular momentum character of the spin-up states (\subfig{\ref{fig:fig3}}{b}) unveils that the combined action of the SOC and orbital hybridization results in the mixing of $d_{z^2}$ states and $m_{-l}=\frac{1}{\sqrt{2}}(d_{x^2-y^2}  -i d_{xy})$ states, in resemblance with the BHZ model of quantum spin Hall insulators, and consequently the edge states of this system carry opposite spin and orbital angular momentum. The quantized spin Hall conductivity characterizes the topological insulator and the opposite bulk spin and orbital Hall conductivities shown in \subfig{\ref{fig:fig3}}{c} confirms the relation with the BHZ model.

\begin{figure}[h]
    \centering
    \includegraphics[width=0.98\linewidth]{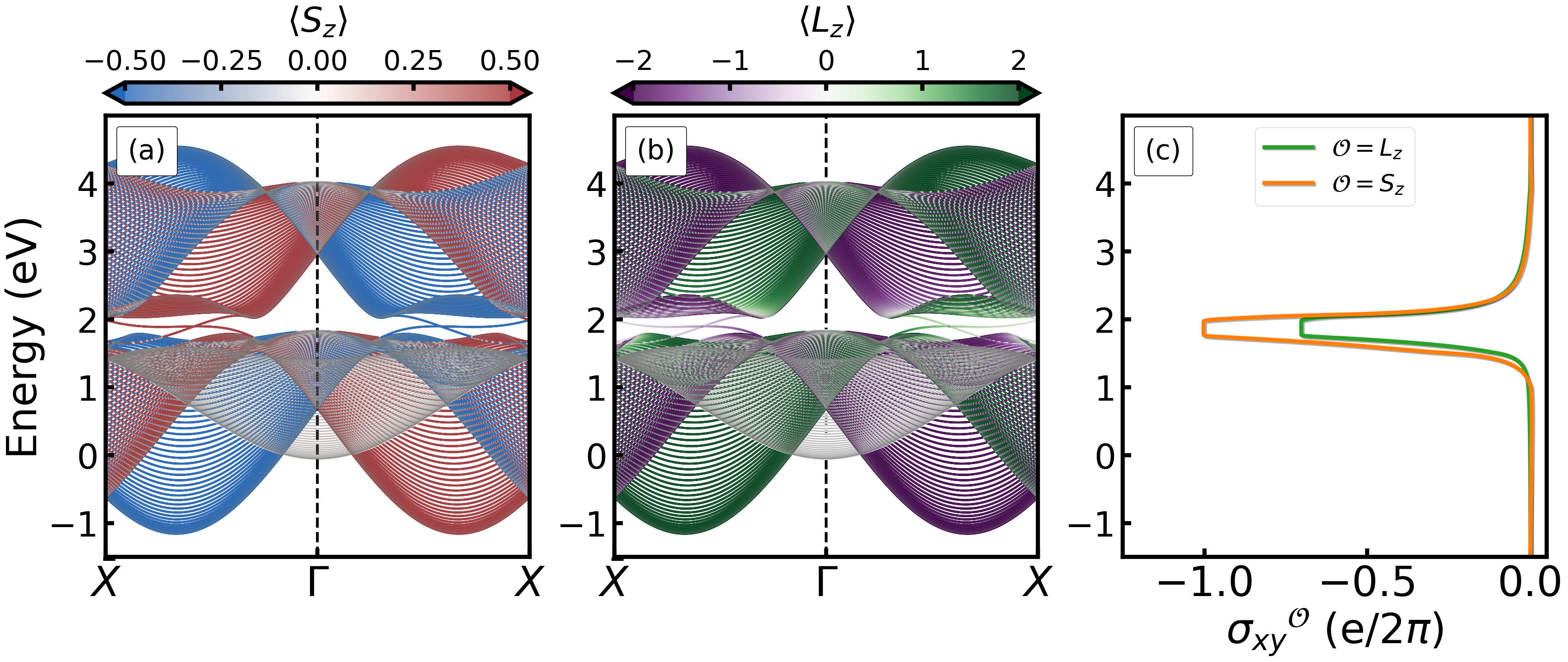}
    \caption{Spin (a) and Orbital (b) character of the energy bands of a nanoribbon with breadth $W=50\vec{a}_2$ for a system in phase III. (c) Spin (orange) and orbital (green) Hall conductivities for a system in phase III. }
    \label{fig:fig4}
\end{figure}

 To further inquire into the effects of the orbital hybridization on the appearance of topologically non-trivial phases, in phase III, we follow a similar analysis as we did for phase II. ./ \subfig{\ref{fig:fig4}}{a} and \subfig{\ref{fig:fig4}}{b} show the band structure and the respective spin and orbital character of the states of the nanoribbon. Here, we forbid the hybridization of the orbitals $d_{z^2}$ with $d_{x^2-y^2},d_{xy}$ by setting the hoppings $t_1=t_2=0$, and we break the inversion symmetry by allowing $t_{12}\neq0$. Analyzing the orbital and spin character of the edge states of the system, we can see that the $d_{x^2-y^2}$ and $d_{xy}$ orbitals are responsible for the non-trivial topological properties of the system, whereas the states related to $d_{z^2}$ are effectively decoupled. In contrast with phase II, the hybridization and the inversion symmetry breaking favor the appearance of pairs of edge states with different spin and orbital characters at opposite sides of the ribbon Brillouin zone, in connection with a spin- and orbital-valley locking. Similarly, as before, these edge states are endowed with sizable orbital angular momentum, providing these states with additional topological protection since any backscattering event would require not only a change of the spin character but also the orbital part. The spin and orbital Hall conductivities shown in \subfig{\ref{fig:fig3}}{c} confirm the topologically non-trivial nature of this phase and the spin and orbital Hall plateaus have the same sign, in contrast with phase II

\begin{figure}[h]
    \centering
    \includegraphics[width=0.98\linewidth]{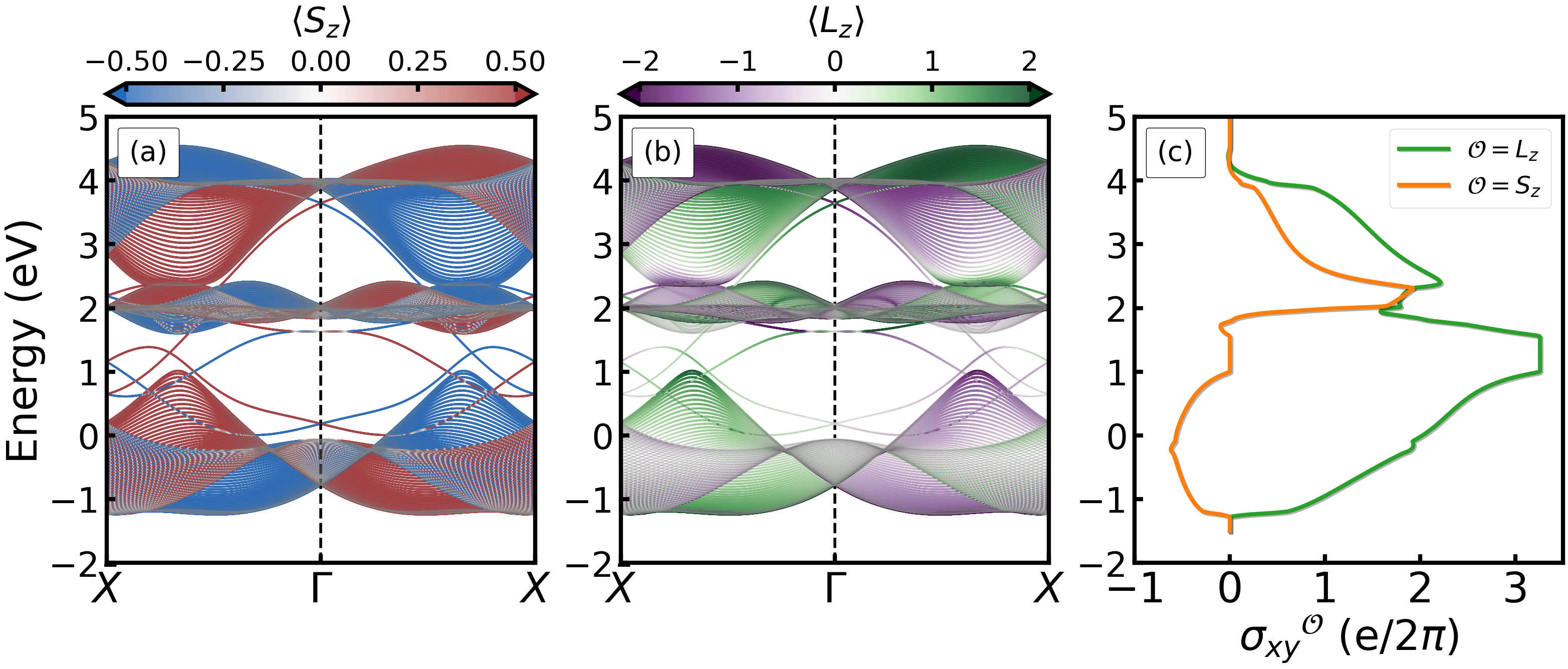}
    \caption{Spin (a) and orbital (b) character of the energy bands of a nanoribbon with breadth $W=50\vec{a}_2$ for a system in phase IV. (c) Spin (orange) and orbital (green) conductivity for a system in phase IV. }
    \label{fig:fig5}
\end{figure}

Finally, for phase IV, we allow all the hopping terms, and essentially the model reduces to the three-band model for 2H-TMDs presented by Liu \textit{et al.} \cite{ThreeBandTMD} but with an enhanced spin-orbit coupling. From analyzing the spin and orbital characters of the energy states of the nanoribbon shown in \subfig{\ref{fig:fig5}}{a} and \subfig{\ref{fig:fig5}}{b}, respectively, we found that the system presents edge states within its main energy gap. They are similar to the edge states presented in MoS$_2$ \cite{Us2,Us3} but with large splitting from the strong spin-orbit coupling. There are four pairs of edge states so that the system is topologically trivial from the $\mathbb{Z}_2$ classification and exhibits a behavior that is drastically different from the previous phases. 
One can notice that the spin character of the edge states in a single side of the ribbon Brillouin zone contains opposite spin polarizations in contrast with the energy states shown in \subfig{\ref{fig:fig4}}{a}. However, when focusing on the orbital character of the edge states of this system, we found that they retain a similar orbital character as shown in \subfig{\ref{fig:fig4}}{b}. These properties correlate with the absence of spin Hall conductivity plateaus within the energy gap but the presence of the orbital Hall insulating phase as shown in \subfig{\ref{fig:fig5}}{c} which is characterized by an orbital Chern number = 1 \cite{Us3}. The even number of edge state pairs in the case of flat termination together with the absence of edge states for zigzag edges is consistent with a higher-order topological phase, which is present in this case \cite{Us4}. 

\section{The orbital Hall conductivity of nanoribbons: Landauer-Büttiker calculations}

\begin{figure}
\includegraphics[scale = 0.3]{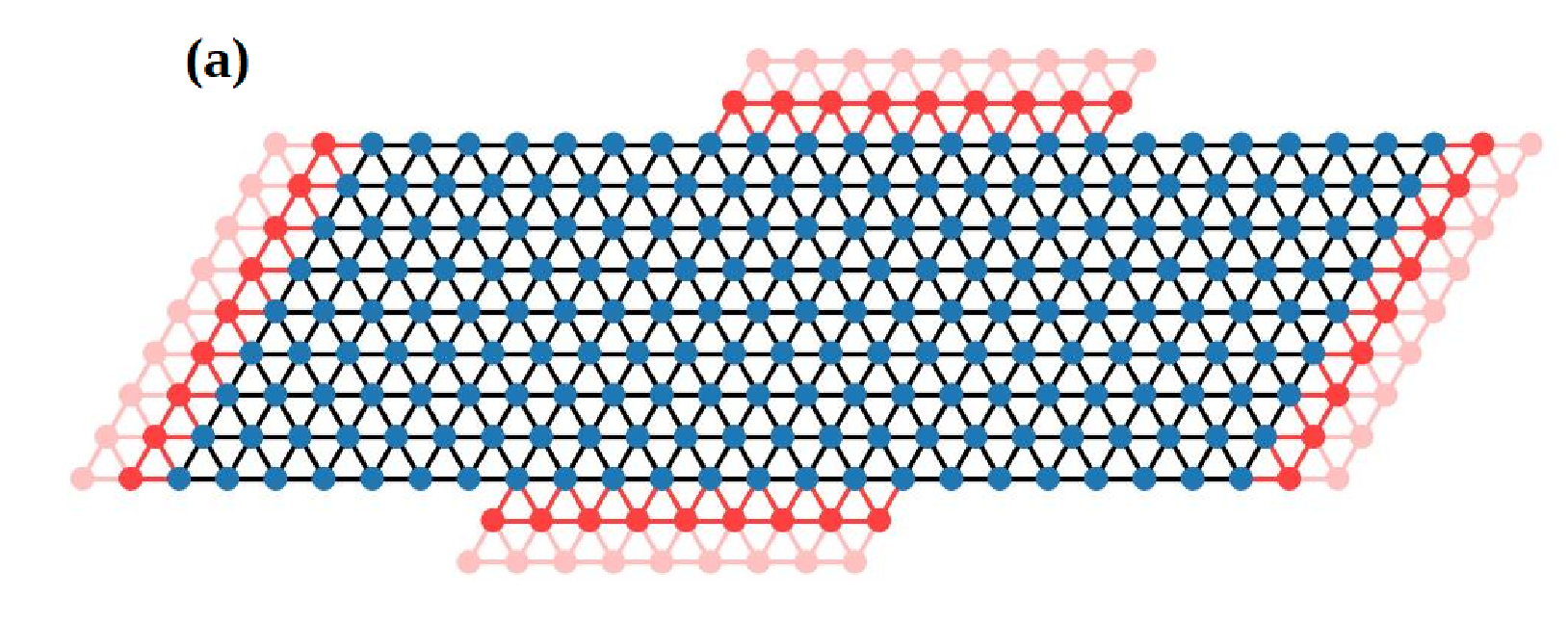}
\includegraphics[scale = 0.55]{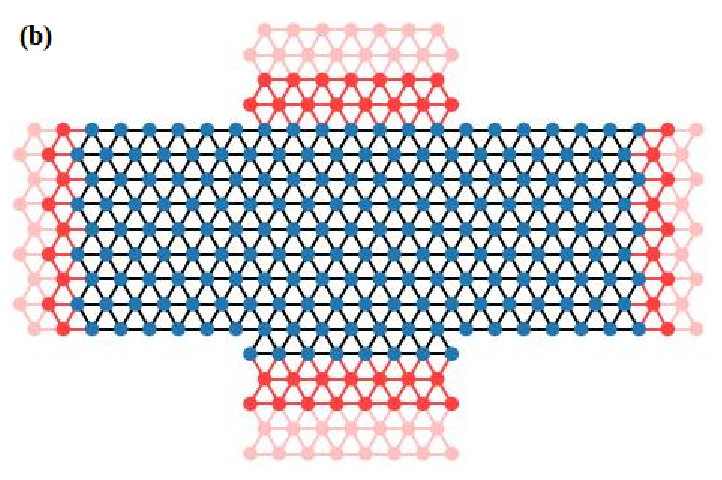}
\caption{(a) Triangular lattice device with edges that either preserve or break the rotational symmetry of the original lattice. (b) Triangular lattice device with edges that break the rotational symmetry of the original lattice. The device (blue) is connected to four leads (red).}\label{sample}
\end{figure}

We proceed to further analyze the higher-order topological phase and check whether the edge states possess the capability to carry orbital angular momentum. To accomplish this, we have designed an OHE setup consisting of a mesoscopic device connected to four semi-infinite terminals subjected to voltage biases $V_i$. In our calculations, we considered two distinct device geometries. \subfig{\ref{sample}}{a} showcases a configuration with all flat edges, whereas \subfig{\ref{sample}}{b} presents a conventional Hall-bar geometry characterized by a combination of flat and zigzag edges.

In the Landauer-B\"uttiker formalism, the orbital (spin) projected current through the $i$th terminal in the linear regime at low temperature is given by
\begin{equation}
    I^{\mathcal{O}}_{i,\eta} = \frac{e^2}{h}\sum_{j} \tau_{ij,\eta}^{\mathcal{O}} \left( V_i - V_j \right),\label{IOS}
\end{equation} 
 with $\mathcal{O}$ being  one of the $L_z$ orbital angular momentum or $S_z$ spin states ($\circlearrowleft(\uparrow)$ or $\circlearrowright(\downarrow)$).
The orbital (spin) transmission coefficient is calculated from the scattering matrix $\mathcal{S}=\left[\mathcal{S}_{ij}\right]_{i,j=1,\dots,4}$ as
\begin{equation}
\tau_{ij,\eta}^{\mathcal{O}} =\textbf{Tr}\left[\left(\mathcal{S}_{ij}\right)^{\dagger} \mathcal{P}^{\mathcal{O}}_{\eta} \mathcal{S}_{ij}\right],
\end{equation}

\noindent where the matrices $\displaystyle \mathcal{P}^{L_\eta}_{\eta} = \mathbb{1}_N \otimes l^\eta \otimes \sigma^0$ and $\mathcal{P}^{S_\eta}_{\eta} = \mathbb{1}_N \otimes l^0 \otimes \sigma^\eta$ are a projector, $\mathbb{1}_N$ is an identity matrix with dimension $N\times N$. The dimensionless integer $N$ refers to the number of propagating wave modes in the terminals, proportional to the terminal width ($W$) and the Fermi vector ($k_F$), computed through the equation $N = k_F W/\pi$.  The matrices $l^{\eta}$ and $\sigma^{\eta}$ with $\eta=\lbrace x,y,z\rbrace$ are the orbital angular momentum and spin matrices, respectively, and the cases with $\eta=0$ refer to the identity matrices in the orbital and spin subspaces. Thus, by setting either $\eta=0$ or $\eta=\{x,y,z\}$, the charge and orbital (spin) can be respectively addressed.


The pure OHC (SHC) $I^{L_z(S_z)}_{i,z}=\frac{\hbar}{e}(I^{\circlearrowleft(\uparrow)}_{i}-I^{\circlearrowright(\downarrow)}_{i}$), $i=3,4$ can be obtained by assuming that the charge current vanishes in the transverse terminals, $I^{c}_{i,0}=I^{\circlearrowleft(\uparrow)}_i+I^{\circlearrowright(\downarrow)}_i=0$, while the charge current is conserved in the longitudinal terminals, $I^c_{1,0}=-I^c_{2,0}=I^c$ \cite{PhysRevB.72.075361,Nikolic_2007,PhysRevLett.98.196601}. By applying these conditions to Eq.(\ref{IOS}), we obtain \cite{fonseca2023orbital}
\begin{eqnarray}
I^{\mathcal{O}}_{i,\eta} = \frac{e}{2\pi}\left[\left(\tau_{i2,\eta}^{\mathcal{O}}-\tau_{i1,\eta}^{\mathcal{O}}\right)\frac{V}{2}
- \tau_{i3,\eta}^{\mathcal{O}}V_3 + \tau_{i4,\eta}^{\mathcal{O}}V_4\right],
\label{Is}
\end{eqnarray}
for $i=3,4$, where $V$ is a constant potential difference between longitudinal terminals, and $V_{3,4}$ is the transversal terminal voltage.

\begin{figure}[h]
    \centering
    \includegraphics[width=0.98\linewidth]{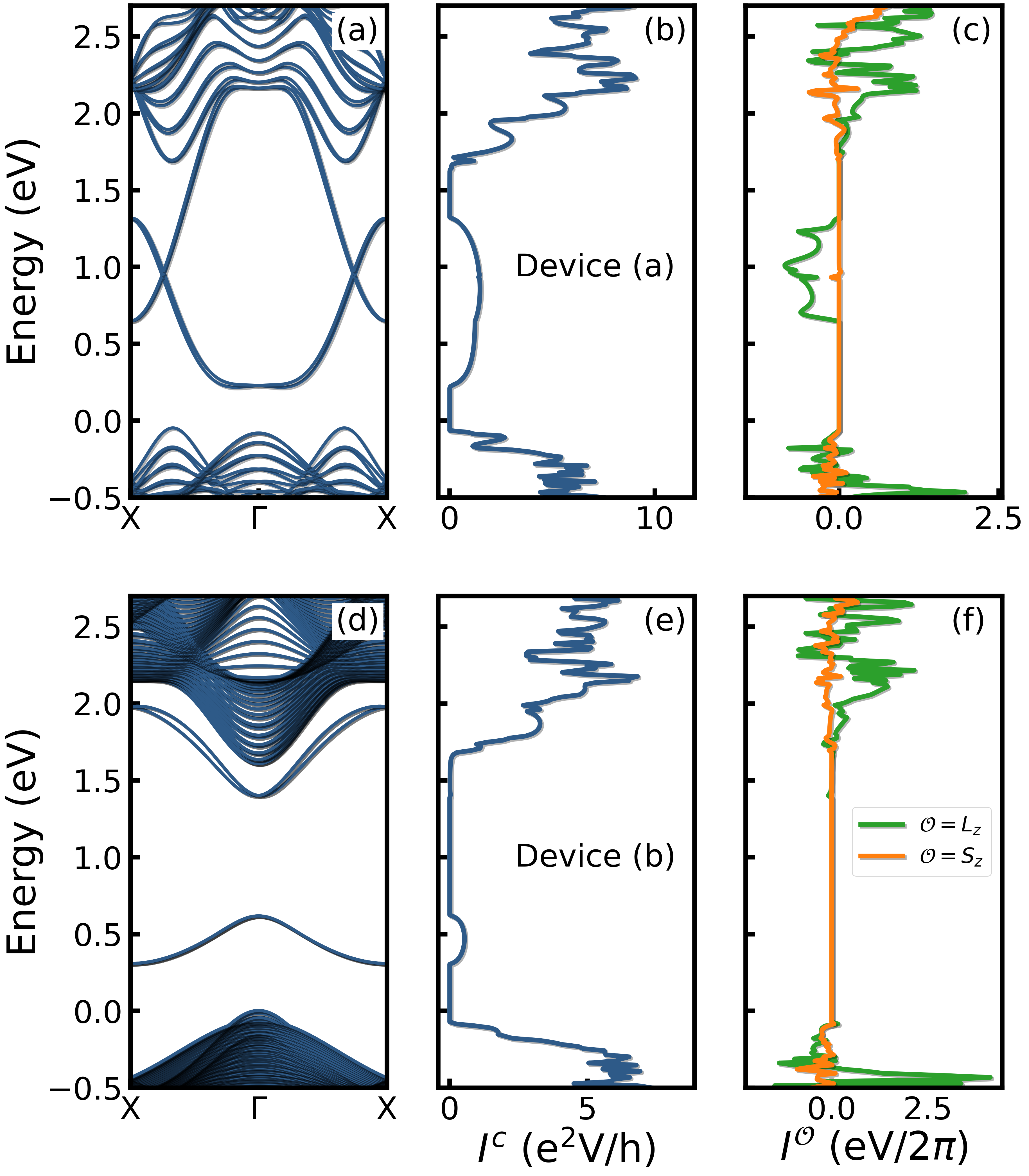}
    \caption{Band structure (a), charge current (b) and orbital and spin Hall current (c) of a mesoscopic device with the dimensions and design of  \subfig{\ref{sample}}{a}. Band structure (d) charge current (e) and orbital and spin Hall current (f) of a mesoscopic device a mesoscopic device with the dimensions and design of \subfig{\ref{sample}}{b}. (c) and (f): Spin (orange) and orbital (green) conductance for a Mo$S_2$ monolayer, which is in phase IV. }
    \label{fig:fig6}
\end{figure}

We begin our investigation by examining the influence of ribbon termination on the orbital Hall conductance in a higher-order topological insulator. To facilitate a clearer visualization of the role of in-gap edge states in orbital transport, we utilize a triangular lattice with parameters akin to MoS$_2$, which belongs to the same topological phase as phase IV. The numerical calculations of this section
were implemented in the KWANT software \cite{Groth_2014}.

It is essential to note that to establish a connection between the edge states of the original honeycomb lattice of MoS$_2$ and the triangular lattice of $d$ orbitals, the zigzag edge of the honeycomb lattice corresponds to the flat edge of the triangular lattice, which consists of one of the sublattices. Simultaneously, the armchair edge of the honeycomb lattice is associated with the zigzag edge of the triangular lattice.

\subfig{\ref{fig:fig6}}{a} presents the band structure of the type of device sketched in \subfig{\ref{sample}}{a} for the parameters of MoS$_2$. In that device, all edges break the three-fold rotation symmetry of the triangular lattice. The band structure of the system is very similar to the band structure of a  MoS$_2$ nanoribbon in the three-orbitals approximation \cite{Costa2023}. \subfig{\ref{fig:fig6}}{b} presents the longitudinal charge current of the device, where it is clear that the in-gap edges are metallic and can conduct charge. \subfig{\ref{fig:fig6}}{c} depicts the transverse spin and orbital currents. While the spin current is zero inside the gap, we can see a finite orbital current in the same energy window of the in-gap edge states. In contrast, if the device also presents some edges that preserve the three-fold rotation symmetry of the triangular lattice, as sketched in figure \subfig{\ref{sample}}{b}, the band structure remains similar, as shown in panel \subfig{\ref{fig:fig6}}{d} but the charge current is suppressed inside the gap (panel e) and the system does not present either spin or orbital currents, as shown in \subfig{\ref{fig:fig6}}{f}.

\begin{figure}[h]
    \centering
    \includegraphics[width=0.98\linewidth]{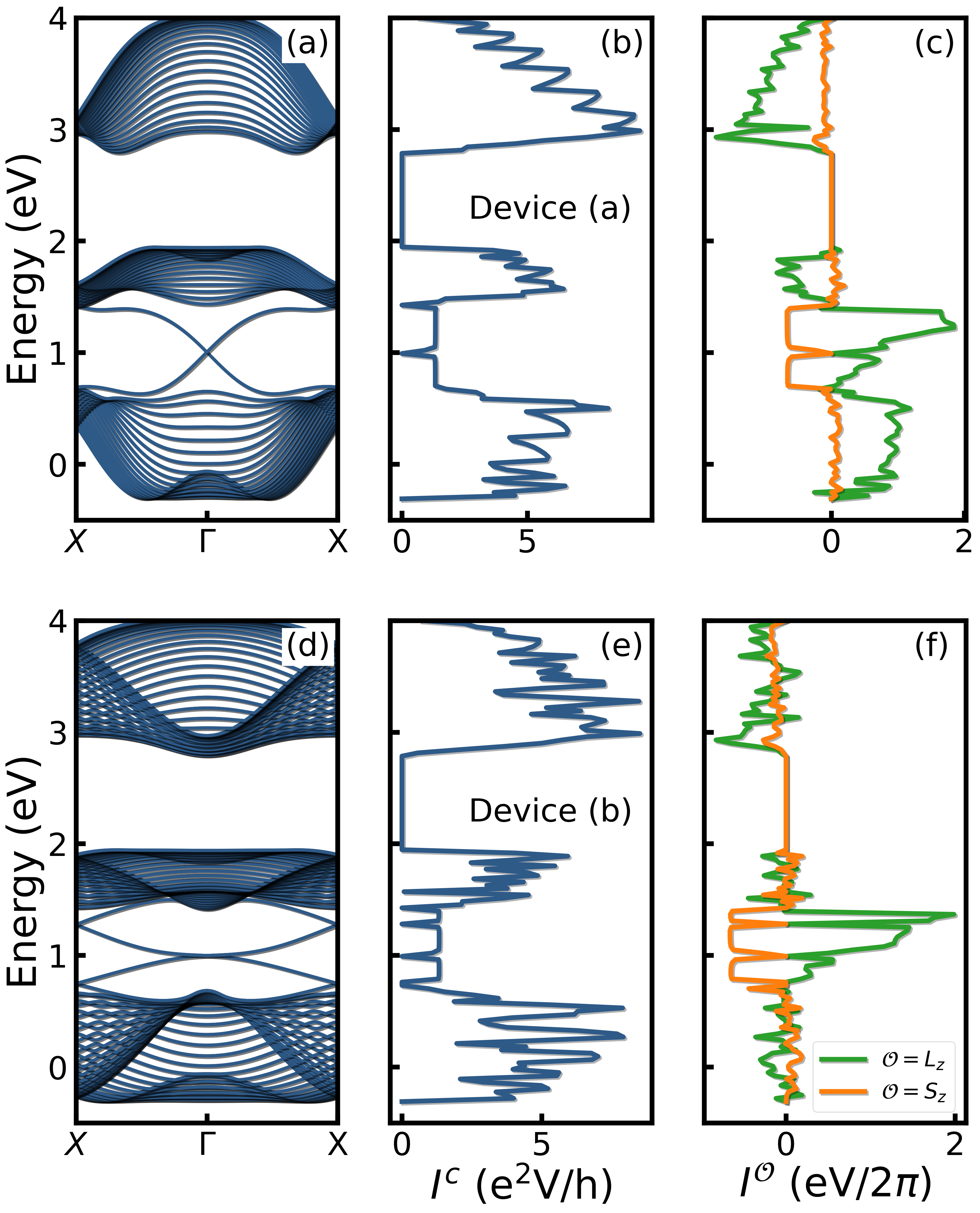}
    \caption{Band structure (a) charge current (b) and orbital and spin Hall current (c) of a mesoscopic device a mesoscopic device with the dimensions and design of \subfig{\ref{sample}}{a}.  Band structure (d) charge current (e) and orbital and spin Hall current (f) of a mesoscopic device a mesoscopic device with the dimensions and design of \subfig{\ref{sample}}{b}. (c) and (f): Spin (orange) and orbital (green) current, which is in phase II. }
    \label{fig:fig7}
\end{figure}

This can be compared with the characteristics of a first-order topological insulator, as the one in phase II.  \subfig{\ref{fig:fig7}}{a} presents the band structure for the device sketched in  \subfig{\ref{sample}}{a} for a topological insulator in phase II, discussed previously. \subfig{\ref{fig:fig7}}{b} presents the longitudinal charge current of the device, where it is clear that the in-gap edges are metallic and can conduct charge. \subfig{\ref{fig:fig7}}{c} depicts the transverse spin and orbital currents. Both spin and orbital currents are finite with opposite signs. While the spin current has a clear plateau inside the gap, the orbital current does not seem to have a quantized value. ./ \ref{fig:fig7}d, e, and f present the same results but for the device sketched in figure \subfig{\ref{sample}}{b}. Differently from the previous cases, the results are robust independent of the type of edges the device presents, underlying the importance of topological protection by a global symmetry.

\begin{figure}[h]
    \centering
    \includegraphics[width=0.98\linewidth]{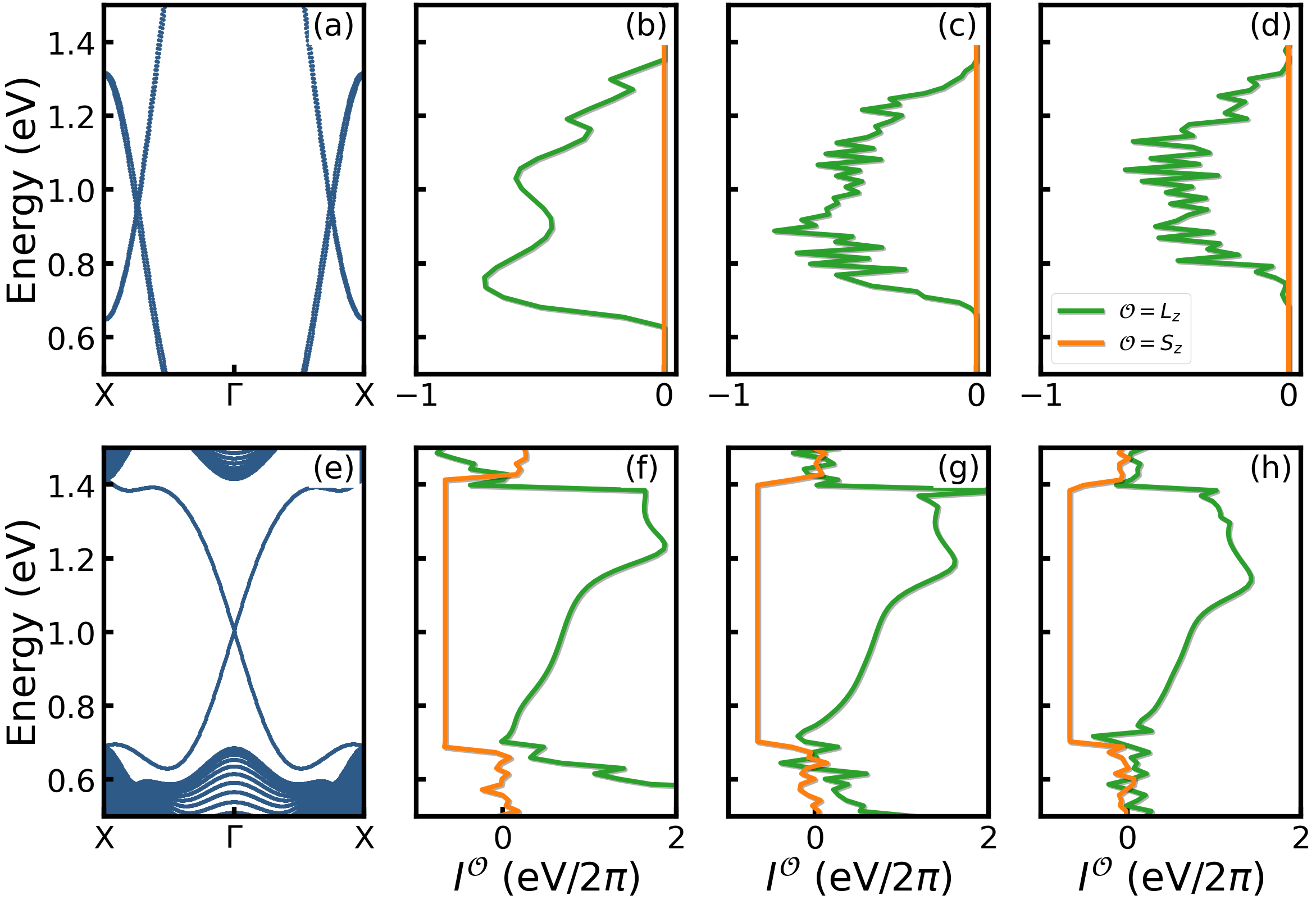}
    \caption{Band structure (a) and orbital and spin Hall current with disorder strengths for $U=0$ (b), $U=0.1$ eV (c), $U=0.2$ eV (d) of a mesoscopic device of Mo$S_2$ monolayer, which is in phase IV, with dimensions ten times bigger than as shown \subfig{\ref{sample}}{a}.  Band structure (e) and orbital and spin Hall current with disorder strengths for $U=0$ (f), $U=0.1$ eV (g), $U=0.2$ eV (h) of a mesoscopic device, which is in phase II, with dimensions ten times bigger than as shown in \subfig{\ref{sample}}{a}. }
    \label{fig:fig8}
\end{figure}

Finally, to study the robustness of the orbital and spin Hall currents against scalar disorder, we included an Anderson disorder term that varies randomly from site to site according to a uniform distribution in the interval $(-U/2,U/2)$, with $U$ being the disorder strength. \subfig{\ref{fig:fig8}}{b}, \subfig{\ref{fig:fig8}}{c}, and \subfig{\ref{fig:fig8}}{d} present both the orbital and spin Hall currents of MoS$_2$ for $U=0$, $U=0.1$ eV, $U=0.2$ eV, respectively. From these ./, one can notice that the orbital Hall currents are not protected against disorder. Although the plateau observed in bulk calculations survives with increasing disorder \cite{PezoDisorder}, our results show a decrease in the orbital Hall current. For the case of the topological insulator of phase II,  panels \subfig{\ref{fig:fig8}}{f}, \subfig{\ref{fig:fig8}}{g}, and \subfig{\ref{fig:fig8}}{h} show that the spin currents are not affected by increment of the disorder, as one would expect for the topologically protected states. In contrast, the orbital Hall currents are affected by the disorder.

\section{Conclusion}
To summarize, our study focuses on a generalized multi-orbital tight-binding model on a triangular lattice. This system was recently used to obtain higher-order topological phases in two-dimensional materials but also is especially relevant for modeling transition metal dichalcogenide monolayers. By modifying the hopping terms of the model, we changed the hybridization between orbitals and uncovered the intricate interplay between spin-orbit coupling and various symmetry-breaking mechanisms, which has led to the identification of four distinct topological phases \cite{Eck2022}. Notably, this interplay has also resulted in the emergence of an orbital Hall effect with unique characteristics. For the two first-order topological phases, the quantized spin Hall conductivity is accompanied by a non-quantized orbital Hall plateau, while the higher-order topological phase presents the orbital Hall effect in the absence of the spin Hall effect.

Furthermore, we have applied the Landauer-B\"uttiker formula to establish that in the orbital Hall insulating phase, the orbital angular momentum is carried by edge states that exist in nanoribbons with flat terminations. Importantly, we have demonstrated that these edge states lack the topological protection against disorder observed in the edge states associated with a first-order topological insulator. These findings provide valuable insights into the behavior of edge states in systems with higher-order topology and their role in the onset of the orbital Hall effect in these systems.

\section{ACKNOWLEDGMENTS}
TGR acknowledges funding from FCT-Portugal through Grant No. CEECIND/07471/2022. L.M.C. acknowledges funding from Ministerio de Ciencia e Innovación de Espa\~na under grant No. PID2019-106684GB-I00 / AEI / 10.13039/501100011033, FJC2021-047300-I, financiado por MCIN/AEI/10.13039/501100011033 and the European Union “NextGenerationEU”/PRTR.". J.H.G. acknowledge funding from MCIN/AEI /10.13039/501100011033 and European Union "NextGenerationEU/PRTR”  under grant PCI2021-122035-2A-2a and funding from the European Union’s Horizon 2020 research and innovation programme under grant agreement No 881603. ICN2 is funded by the CERCA Programme/Generalitat de Catalunya and supported by the Severo Ochoa Centres of Excellence programme, Grant CEX2021-001214-S, funded by MCIN/AEI/10.13039.501100011033. ALRB acknowledges financial support from Conselho Nacional de Desenvolvimento Cient\'{\i}fico e Tecnol\'ogico - CNPQ (Grant No 309457/2021-1).

%

\onecolumngrid

\appendix
\section{Slater-Koster for $p$ orbitals in a triangular lattice}
To calculate the first-neighbors hoppings of a hamiltonian of $p$ orbitals in a triangular lattice, we define $n_z=\cos{\theta}$, $n_x=\cos{\phi}\sin{\theta}$ and $n_y=\sin{\phi}\sin{\theta}$.

For the hopping between $p$ orbitals one has
\begin{align}
t_{p_i,p_i}=n_i^2V_{pp\sigma}+(1-n_i^2)V_{pp\pi} \\
t_{p_i,p_j}=n_in_j(V_{pp\sigma}-V_{pp\pi})
\end{align}
with $i = x,y,z$ and $i\neq j$.

For two-dimensional lattice, we define $n_z=\cos{\theta=0}=1$, $n_x=\cos{\phi}$ and $n_y=\sin{\phi}$

and  $t_{p_z,p_z}=n_i^2V^z_{pp\sigma}$.
Using $\cos(\pi/3)=\cos(-\pi/3)=-\cos(2\pi/3)=-\cos(-2\pi/3)=\frac{1}{2}$, $\sin(\pi/3)=-\cos(-\pi/3)=\sin(2\pi/3)=-\sin(-2\pi/3)=\frac{\sqrt{3}}{2}$, $\cos(0)=-\cos(\pi)=1$ and $\sin{0}=\sin(\pi)=0$ 

one can construct the hopping elements that preserve all the symmetries of the lattice

\begin{align}
h_{p_x,p_x}(k_x,k_y)&=\frac{1}{4}(V_{pp\sigma}+3V_{pp\pi})(e^{i(\ka+\kb)}+e^{-i(\ka+\kb)a}+e^{i(\ka-\kb)}+e^{-i(\ka-\kb)})\\
&+V_{pp\sigma}(e^{2i\ka}+e^{-2i\ka})\\
h_{p_x,p_x}(k_x,k_y)&=(V_{pp\sigma}+3V_{pp\pi})\cos\ka \cos\kb +2V_{pp\sigma}\cos{2\ka}\\
h_{p_y,p_y}(k_x,k_y)&=(3V_{pp\sigma}+V_{pp\pi})\cos\ka \cos\kb +2V_{pp\pi}\cos{2\ka}
\end{align}

using the expressions above, one has  $\cos(\pi/3)\sin(\pi/3)=-\cos(-\pi/3)\sin(-\pi/3)=\cos(-2\pi/3)\sin(-2\pi/3)=-\cos(2\pi/3)\sin(2\pi/3)=\frac{\sqrt{3}}{4}$ and $\cos(0)\sin(0)=\cos(\pi)\sin(\pi)=0$
\begin{align}
h_{p_x,p_y}(k_x,k_y)&=\frac{\sqrt{3}}{4}(V_{pp\sigma}-V_{pp\pi})(e^{i(\ka+\kb)}+e^{-i(\ka+\kb)}-e^{i(\ka-\kb)}-e^{-i(\ka-\kb)}) \\
h_{p_x,p_y}(k_x,k_y)&=-\sqrt{3}(V_{pp\sigma}-V_{pp\pi})\sin \ka \sin \kb.
\end{align}
As can be seen, if all the symmetries are preserved, the subspace of the $p_z$ orbital is separated from the subspace of $p_x$ and $p_y$ orbitals. However, if the vertical reflection symmetry is broken, it allows for the  hybridization between them, which can be given by $t_{p_ip_z}=\lambda_hn_i=-t_{p_zp_i}$
\begin{align}
 h_{p_x,p_z}(k_x,k_y)=&\frac{1}{2}\lambda_h(e^{i(\ka+\kb)}-e^{-i(\ka+\kb)}+e^{i(\ka-\kb)}-e^{-i(\ka-\kb)}+2e^{2i\ka}-2e^{-2i\ka})\\
 h_{p_x,p_z}(k_x,k_y)=&-2i\lambda_h(\sin{2\ka}+\sin\ka \cos\kb)\\
 h_{p_y,p_z}(k_x,k_y)=&\frac{\sqrt{3}}{2}\lambda_h(e^{i(\ka+\kb)}-e^{-i(\ka+\kb)}-e^{i(\ka-\kb)}+e^{-i(\ka-\kb)})\\
 h_{p_y,p_z}(k_x,k_y)=&2\sqrt{3}i\lambda_h\cos\ka\sin\kb.
\end{align}
We can also break inversion symmetry by breaking $\sigma_v$ while still respecting the three-fold rotation symmetry. For this purpose, we can introduce a hopping given by   $t^v_{p_y,p_x}=\lambda_v\cos(3\phi)=-t^v_{p_x,p_y}$, so that
\begin{align}
 h^v_{p_x,p_y}(k_x,k_y)=&\lambda_v(-e^{i(\ka+\kb)}+e^{-i(\ka+\kb)}-e^{i(\ka-\kb)}+e^{-i(\ka-\kb)}+e^{2i\ka}-e^{-2i\ka})\\
 h^v_{p_x,p_y}(k_x,k_y)=&\lambda_v(2i\sin{2\ka}-4i\sin\ka \cos\kb)\\
  h^v_{p_x,p_y}(k_x,k_y)=&4i\lambda_v\sin{\ka}(\cos\ka -\cos\kb).
\end{align}
\section{Low Energy Hamiltonian}

To better understand how phase IV is related to the appearance of an orbital Hall plateau in the absence of spin Hall conductivity, it is helpful to look at a simplified low-energy Hamiltonian. Since the original Hamiltonian, $H_0$, is written in a basis that spans states with orbital angular momentum quantum numbers $m_l$ of $-l$, $0$, and $l$, we can map it to a basis of $p$ orbitals, and rewrite it in terms of the orbital angular momentum operators $L_x$, $L_y$, and $L_z$. We can then expand this Hamiltonian around the Dirac points, which is relevant for phase IV:

The Hamiltonian for each valley $H_{K(K^\prime)}(q_x,q_y)$, with $\vec{q}=\vec{k}-\vec{K}(\vec{K}^\prime)$ reads as 
\begin{widetext}
\begin{align}
H_K(q_x,q_y)=&\left(-\frac{3}{2}(t_{11}+t_{22})+\epsilon_2\right)L_z^2-3\sqrt{3}t_{12}\tau_z L_z+\lambda s_z L_z+(-3 t_0 + \epsilon_1)\left(\frac{L^2}{l(l+1)}-L_z^2\right)+\frac{3\sqrt{3}}{2}t_2\tau_z(q_xL_x-q_yL_y)\nonumber\\
&+\frac{3}{2}t_1\left\{L_z,(q_xL_x-q_yL_y)\right\}-\frac{3\sqrt{3}}{4}(t_{11}-t_{22})\left(L_x(q_xL_x+q_yL_y)+L_y(q_xL_y-q_yL_x)\right)
\end{align}
\end{widetext}
where $\tau_z=\pm 1$ represents the different valleys and $s_z$ is the spin operator. From the equation above, it is clear that up to first order in $q_x$ and $q_y$ the spin-orbit coupling does not produce any spin texture. On the other hand, the last term of the Hamiltonian, proportional to the difference $t_{11}-t_{22}$, is connected to an orbital Rashba coupling while the terms proportional to $t_1$ and $t_2$ incorporate an orbital Dresselhaus coupling. Furthermore, $t_{12}$ is responsible for a strong orbital-valley locking when it dominates over other contributions.  It becomes clear that the spin-valley coupling that is seeing in TMDs is mediated by an orbital-valley coupling

\begin{figure}[h]
    \centering
    \includegraphics[width=0.98\linewidth]{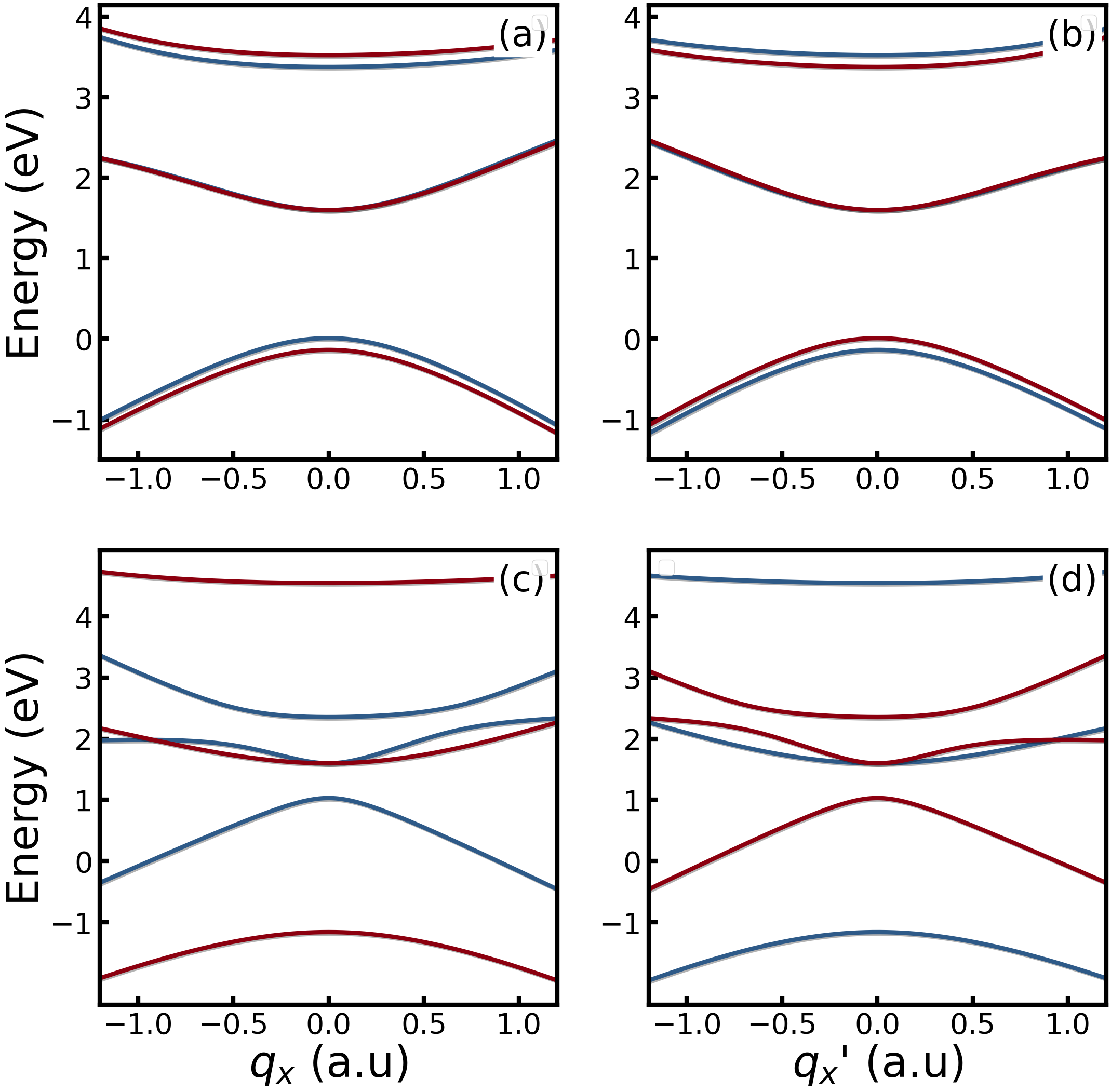}
    \caption{Spin (a) and orbital (b) character of the energy bands of a nanoribbon with breadth $W=50\vec{a}_2$ for a system in the phase IV. (c) Spin (orange) and orbital (green) conductivity for a system in the phase IV. }
    \label{fig:figAp}
\end{figure}

To test the new low-energy Hamiltonian, we first present the band structure for the parameters of the three bands model of MoS$_2$ in figure \ref{fig:figAp} in the vicinity of $K$ (a) and $K^\prime$ Dirac points (b). The band structure is consistent with both the tight-binding and the two-bands low energy model, keeping the spin character of the bands in the two Dirac points while explicitly taking into account the orbital characters of the three states in $K$ and $K^\prime$. figure \ref{fig:figAp}c-d show the band structure for the parameters used in Phase IV and it is able to also provide a consistent description of the bands.

\end{document}